\documentstyle[14pt]{article}

\begin{document}           
\title{
 Report IRB-StP-GR-010896 \newline
\newline \newline \newline
{\bf
An Exact Solution with $f^2 = 1$ and $\Lambda \ne 0$ in the LTB model.
}
}
\author{\it by \\ \\
{\bf Alexander Gromov}
\\ \\
\small\it St. Petersburg State Technical University \\
\small\it Faculty of Technical Cybernetics, Dept. of Computer Science \\
\small\it 29, Polytechnicheskaya str. St.-Petersburg, 195251, Russia \\
\small and \\
\small\it Istituto per la Ricerca di Base \\
\small\it Castello Principe Pignatelli del Comune di Monteroduni \\
\small\it I-86075 Monteroduni(IS), Molise, Italia \\
\small\it e-mail: gromov@natus.stud.pu.ru
}
\date{}
\maketitle
\begin{abstract}

The exact solution in the LTB model with $f^2 = 1$, $\Lambda \ne 0$ is
studied. The initial conditions for the metrical function and its
derivatives generate the solution with complicated structure including
the solutions like "stripping of the shell", "collapce" and "core", or
"accretion". In the limit of big time the solution allows the constant
Hubble function and the density, depending on time.
The transformation to the FRW model is shown.
\\
PACS number(s):98.80
\end{abstract}
\newpage
\section{The Introduction} \label{introd}

As it is shown in \cite{A.G.1}
the LTB model \cite{Tolman:34} is reduced to the Cauchy problem for the
equation
\begin{equation}
\left(
\frac{\partial e^{\omega(\mu,\tau)/2}}{\partial \tau}
\right)^2 =
f^2(\mu) - 1 + \frac{1}{2} F(\mu) e^{-\omega(\mu,\tau)/2} +
\frac{\Lambda}{3} e^{\omega}
\label{dif_1}
\end{equation}
whit initial conditions \cite{A.G.1}
\begin{equation}
\left.
\begin{array}{c}
\left.\omega(\mu,\tau)\right|_{\tau=0} = \omega_0(\mu), \quad
\left.\dot\omega(\mu,\tau)\right|_{\tau=0} = \dot\omega_0(\mu), \quad \\ \\
\left.\ddot\omega(\mu,\tau)\right|_{\tau=0} = \ddot\omega_0(\mu)
\end{array}
\right\}
\label{init}
\end{equation}
and constants
\begin{eqnarray}
\left.
\begin{array}{c}
\left.\omega(\mu,0)\right|_{\mu=0} = \omega_0(0), \quad
\left.\dot\omega(\mu,0)\right|_{\mu=0} = \dot\omega_0(0), \quad \\ \\
\left.\ddot\omega(\mu,0)\right|_{\mu=0} = \ddot\omega_0(0), \quad
\Lambda,
\end{array}
\right\}
\label{init-c}
\end{eqnarray}
where $\mu$ and $\tau$ are dimensionless co-moving coordinates,
corresponding to Lagrangian coordinate and time,
$\omega(\mu,\tau)$ is the metrical function,
$\dot{} = \frac{\partial}{\partial \tau}$, $\Lambda$ is the cosmological
constant,
\begin{equation}
f^2(\mu) - 1 =
e^{\omega_0(\mu)}
\left(
\ddot \omega_0(\mu) + \frac{3}{4} \dot \omega^2_0(\mu) - \Lambda
\right),
\label{DEF:f^2-1}
\end{equation}
\begin{equation}
F{\rm(\mu)} =
e^{3 \,\omega_0(\mu) / 2}
\left(
\frac{\dot \omega^2_0(\mu)}{2} - \frac{2}{3} \Lambda
\right)
+2
e^{\omega_0(\mu) /2}
\left[1 - f^2(\mu)\right].
\label{DEF:itF}
\end{equation}
In this article we study the class of the LTB models defined by the
follow of initial conditions:
\begin{equation}
f^2(\mu) = 1, \qquad \Lambda \ne 0.
\label{in-cond}
\end{equation}
The Bonnor function \cite{Bonnor:72}
\begin{equation}
R(\mu,\tau) = e^{\omega(\mu,\tau)/2}
\label{Bonnor}
\end{equation}
will be also used.
For the obtained solution will be studied the Hubble function
\begin{equation}
h(\mu,\tau) = \frac{\dot R^{\prime}(\mu,\tau)}{R^{\prime}(\mu,\tau)}
= \frac{\partial \ln{R^{\prime}(\mu,\tau)}}{\partial \tau}
\label{Hubble_def}
\end{equation}
and the density law \cite{A.G.2}
\begin{eqnarray}
{\rm
    8 \pi \delta(\mu,\tau) =
    \frac{e^{
             \frac{3}{2}
             [\omega_0(\mu) - \omega(\mu,\tau)]
            }
         }
         {
          \omega^{\prime}(\mu,\tau)
         }
} \times
\nonumber\\
{\rm
    \left\{
           3\left[\omega_0(\mu)\right]^{\prime}
            \left[
                  -\ddot\omega_0(\mu) - \frac{1}{2}\dot\omega_0^2(\mu) +
                  \frac{\Lambda}{6}
            \right]
     - 2\left[\ddot\omega_0(\mu)\right]^{\prime} -
     2\dot\omega_0(\mu)\left[\dot\omega_0(\mu)\right]^{\prime}
    \right\}
}
\label{DEF:rho_new}
\end{eqnarray}
in the limit of $\tau \rightarrow +\infty$.

\section{The Input Equation and Previous Analysis} \label{solution-1}

The initial conditions (\ref{in-cond}) produce the equation which we are
interested in:
\begin{equation}
\frac{3}{\Lambda}\,r(\mu,\tau)\,
\left(\frac{\partial r(\mu,\tau)}{\partial \tau}\right)^2 =
k(\mu) + r^3(\mu,\tau),
\label{dif_2}
\end{equation}
where
\begin{equation}
r(\mu,\tau) = \frac{R(\mu,\tau)}{R_0(\mu)}, \qquad R_0(\mu) =
\left.R(\mu,\tau)\right|_{\tau = 0},
\label{def r}
\end{equation}
\begin{equation}
k(\mu) = \frac{3}{4}\frac{\dot\omega^2_0(\mu)}{\Lambda} - 1.
\label{sign}
\end{equation}
The initial condition for the equation (\ref{dif_2}) is
\begin{equation}
r(\mu,0) = 1.
\label{r-ini}
\end{equation}
In accordance with (\ref{DEF:f^2-1}),
the general number of the initial conditions (\ref{init}) is decreased by
one. (\ref{in-cond}) binds two initial conditions $\omega_0(\mu)$
and $\dot\omega_0(\mu)$ as well:
\begin{equation}
\ddot \omega_0(\mu) + \frac{3}{4} \dot \omega^2_0(\mu) - \Lambda = 0.
\label{in-cond-m}
\end{equation}
The definition of the initial value of the velocity
follows from (\ref{dif_2}) and (\ref{r-ini}):
\begin{equation}
\frac{3}{\Lambda}\,
\left.
\left(\frac{\partial r(\mu,\tau)}{\partial \tau}\right)^2
\right|_{\tau = 0}
= k(\mu) + 1,
\label{ini-v}
\end{equation}
and, so, the restriction for function $k(\mu)$ has the form:
\begin{equation}
k(\mu) \ge - 1.
\label{vv}
\end{equation}
The boundary value $k(\mu) = -1$ is reached in two cases:
1) $\dot\omega_0(\mu) = 0$,
in this case the initial velocity is equal to zero:
\begin{equation}
\left.\frac{\partial r(\mu,\tau)}{\partial \tau}\right|_{\tau = 0}
 = 0;
\label{ini-v-zero}
\end{equation}
or 2) under the approximation
"$\Lambda \rightarrow +\infty$" and the initial velocity
being an arbitrary finite function.

It follows also from (\ref{dif_2}) that the minimal distance from the
centre and the particle is
\begin{equation}
r_{min}(\mu,\tau) = |k(\mu)|^{1/3}.
\label{m - d}
\end{equation}

Before solving the equation (\ref{dif_2}) we will study two limit cases
of this equation.
They appear in the process of the competition of two summands
at the right part of the equation (\ref{dif_2}):
$k(\mu)$ and $r^3(\mu,\tau)$. Equating them, we obtain the
definition of the boundary, presented by the equation (\ref{m - d}).
First limit case is that of small $r(\mu,\tau)$:
\begin{equation}
0 < r(\mu,\tau) \ll |k(\mu)|^{1/3}.
\label{1-asimpt - <<f}
\end{equation}
In this case the main contribution to the exact solution, satisfying the
initial condition (\ref{r-ini}), is given by the formula
\begin{equation}
r^{main}_{<<}(\mu,\tau) = \left(
1 \pm \sqrt{
|k(\mu)|}\frac{\tau}{\tau_0}
\right)^{2/3},
\label{2-asimpt - <<}
\end{equation}
where
\begin{equation}
\tau_0 = \frac{2}{\sqrt{3 \Lambda}}.
\label{tau_00}
\end{equation}
This solution has the form of the Bonnor's solution \cite{Bonnor:72}.

The second limit case is that of big $r(\mu,\tau)$:
\begin{equation}
r(\mu,\tau) \gg |k(\mu)|^{1/3}
\label{1-asimpt - >>}
\end{equation}
The main contribution to the exact solution in this case
is given by the formula
\begin{equation}
r^{main}_{>>}(\mu,\tau)  = e^{\pm\frac{2}{3}\frac{\tau}{\tau_0}}.
\label{2-asimpt - >>}
\end{equation}
The initial conditions are "forgotten" in this limit and the dynamics is
defined only by $\Lambda$.

Let us study the input equation in the limit $\Lambda \rightarrow 0$.
The equation (\ref{dif_2}) reads
\begin{equation}
3\,r(\mu,\tau)\,
\left(\frac{\partial r(\mu,\tau)}{\partial \tau}\right)^2 =
\frac{3}{4}\dot\omega^2_0(\mu) + \Lambda \left(r^3(\mu,\tau) - 1\right)
\label{dif_2-limit}
\end{equation}
and reduced in this limit to the equation
\begin{equation}
\frac{\partial r^{3/2}(\mu,\tau)}{\partial \tau} =
\pm\,\frac{3}{4}\dot\omega_0(\mu).
\label{dif_2-limit-a}
\end{equation}
This equation is studied in \cite{A.G.2}. It is shown there that for
$\dot\omega_0(\mu) = const$ the equation (\ref{dif_2-limit-a}) reduced to
the FRW model.
In case $\dot\omega_0(\mu) \ne const$ the equation (\ref{dif_2-limit-a})
includes the FRW model as the main part of the exact solution
in the limit of big time: $\tau \gg \frac{4}{3\,\dot\omega_0(\mu)}$.

We will study now the exact solution of the equation (\ref{dif_2}).
The equation depends on the sign of $k(\mu)$
and its solution is reduced to the calculation of the following integral:
\begin{equation}
\frac{1}{\sqrt{|k(\mu)|}}\,
\int
\frac{d\left(r^{3/2}\right)}
     {\sqrt
           {\frac{1}{|k(\mu)|}r^3 + sign\,k(\mu)}
     } =
{\bf F}(\mu) \pm \frac{\tau}{\tau_0},
\label{int:}
\end{equation}
the function ${\bf F}(\mu)$ being an undetermined function will be defined
by the initial conditions as follows:
\begin{equation}
{\bf F}(\mu) =
\frac{1}{\sqrt{|k(\mu)|}}\,
\left.\int
\frac{d\left(r^{3/2}\right)}
     {
     \sqrt
         {
          \frac
           {1}
           {|k(\mu)|} r^3
           + sign\,k(\mu)
         }
}\right|_{\tau = 0}.
\label{def F}
\end{equation}
The upper sign in the equation (\ref{int:})
corresponds to the expansion from the initial state
into the infinity and the lower sign corresponds to the fall to
the its centre. We will name the solution with sign $+$ the "expanding"
solution and the solution with $-$ the "falling" solution.

\section{The "pure" solutions (function $k(\mu)$ has a constant sign)}
\label{solution-2}

We will study three cases: $sign\, k(\mu) = -1$, $sign\,k(\mu) = 0$,
$sign\,k(\mu)  = 1$.

Case 1: $sign\,k(\mu) = -1$. The equation (\ref{int:}) reads:
\begin{equation}
\frac{1}{\sqrt{|k(\mu)|}}\,
\int
\frac{d\left(r^{3/2}\right)}
     {\sqrt
           {\frac{1}{|k(\mu)|} r^3 -1}
     } =
{\bf F}_{1}(\mu) \pm \tau,
\label{int:case 1}
\end{equation}
and after the calculation of the integral has the form
\begin{equation}
r^{3/2}(\mu,\tau) =
\sqrt{|k(\mu)|}\,
ch\left(
{\bf F}_{1}(\mu)
\,\pm \frac{\tau}{\tau_0}
\right),
\label{res case 1}
\end{equation}
where
\begin{equation}
{\bf F}_{1}(\mu)
= Arch\frac{1}{\sqrt{|k(\mu)|}} = Arsh\left(
\frac{2}{3} \frac{\tau_0}{\sqrt{|k(\mu)|}} \dot r(\mu,0)
\right)
,
\label{def tau_0}
\end{equation}

First we point out the special case $k(\mu) = -1$. Its peculiarity
is the following: at the moment of time $\tau = 0$
all particles have the velocity equal to zero: ${\bf F}_{1}(\mu) = 0$.
The particles spread out to infinity from the point $r(\mu,0) = 1$.

We will study now the falling solution more detail.
All particles start from the initial position $r(\mu,0) = 1$.
In accordance with the solution (\ref{res case 1}) the velocity of the
particle $\frac{\partial r(\mu,\tau)}{\partial \tau}$ is equal to zero at
time
\begin{equation}
\bar\tau(\mu) = \tau_0 {\bf F}_{1}(\mu)\,.
\label{zero velocity}
\end{equation}
The distance from the centre to the particle at this moment is
\begin{equation}
r(\mu,\bar\tau) =
|k(\mu)|^{1/3}.
\label{minimum distance}
\end{equation}

There are no particles which are able to fall onto the centre $r = 0$
Because in this case $|k(\mu)| \ne 0$. So, all particles with Lagrangian
coordinates from the set
\begin{equation}
0 < \dot\omega^2_0(\mu) < \frac{4}{3}\Lambda
\label{case 1 peculiarity}
\end{equation}
fall to the centre at the period of time
$0 < \tau < \bar\tau(\mu)$, have the zero velocity
at time $\tau = \bar\tau$ and go to the infinity after $\bar\tau$.
Because of this Case 1 named "stripping of the shell".

Case 2: $k(\mu) = 0$. The equation (\ref{dif_2}) reads:
\begin{equation}
\frac{\partial r}{\partial \tau} = \pm \sqrt{\frac{\Lambda}{3}} r =
\pm \frac{2}{3} \frac{\tau}{\tau_0}
\label{case 3}
\end{equation}
and has the solution
\begin{equation}
r(\mu,\tau) = e^{\pm \frac{2}{3}\frac{\tau}{\tau_0}}, \qquad {\bf
F}_{2}(\mu) = 0.
\label{sol case-3}
\end{equation}
The expanded solution has no peculiarity. The falling solution shows that
the particles with Lagrangian coordinate
\begin{equation}
\dot\omega^2_0(\mu) = \frac{4}{3}\Lambda
\label{case 2 peculiarity}
\end{equation}
come to the centre for infinite time with zero velocity.
Because of this Case 2 named "collapse".

This case coincides with the limit (\ref{1-asimpt - >>}) of the equation
(\ref{dif_2}).

Case 3: $sign\,k(\mu) = 1$. The equation (\ref{int:}) reads:
\begin{equation}
\frac{1}{\sqrt{k(\mu)}}\,
\int
\frac{d\left(r^{3/2}\right)}
     {\sqrt
           {\frac{1}{k(\mu)} r^3 + 1}
     } =
{\bf F}_{3}(\mu) \pm \tau,
\label{int:case 3}
\end{equation}
and after calculation of the integral has the form
\begin{equation}
r^{3/2}(\mu,\tau) =
\sqrt{k(\mu)}\,
sh\left(
{\bf F}_{3}(\mu) \,
\pm \frac{\tau}{\tau_0}
\right),
\label{res case 3}
\end{equation}
where
\begin{equation}
{\bf F}_{3}(\mu)\, = Arsh\frac{1}{\sqrt{k(\mu)}},
\label{def tau_0 3}
\end{equation}
In accordance with the solution (\ref{res case 3}) the velocity
$\frac{\partial r(\mu,\tau)}{\partial \tau}$
of the particle with Lagrangian coordinates satisfied the condition
\begin{equation}
\dot\omega^2_0(\mu) > \frac{4}{3}\Lambda
\label{case 3 peculiarity}
\end{equation}
is never equal to zero and these particles reach the centre at the moment
of time
\begin{equation}
\bar\tau(\mu) = \tau_0 \, {\bf F}_{3}(\mu).
\label{zero velocity 3}
\end{equation}
Because of this Case 3 case named "accretion".

We will study now the condition of application of the Bonnor's solution
(\ref{2-asimpt - <<}) which takes a place in the limit (\ref{2-asimpt -
<<}). First, we rewrite the solution (\ref{int:case 3}) in the form:
\begin{equation}
r^{3/2}(\mu,\tau) = ch\frac{\tau}{\tau_0} \pm \sqrt{k_3(\mu) +
1}\,sh\frac{\tau}{\tau_0}.
\label{c:3}
\end{equation}
We will transform this solution, taking in ... the fact that
\begin{equation}
k_3(\mu) \gg 1.
\label{c:4}
\end{equation}
It is follows from (\ref{c:3}) in this case
\begin{equation}
Arsh\frac{1}{\sqrt{k_3(\mu)}} = {\bf F}_3(\mu) =
\frac{\bar\tau(\mu)}{\tau_0} < 1.
\label{c:31}
\end{equation}
For
\begin{equation}
\tau < \bar\tau < \tau_0
\label{c:42}
\end{equation}
it is follows from (\ref{c:3}) - (\ref{c:31}) that
\begin{equation}
r^{3/2}(\mu,\tau) \approx 1 \pm\, \sqrt{k_3(\mu)}\,\frac{\tau}{\tau_0},
\label{c:41}
\end{equation}
what equivalent to the equation (\ref{2-asimpt - <<}).
All particles with Lagrangian coordinates satisfied the condition
(\ref{c:4}), form the core with equation of motion containing the Bonnor's
solution as the main part. So, the Case 3 contains the Bonnor's core and
nonbonnor's part.

\section{The Asymptotic Propertys of the Solution} \label{solution-3}

In this section the density law (\ref{DEF:rho_new})
and the Hubble function (\ref{Hubble_def}) will be studied
for the expanding solution with in the limit
\begin{equation}
\tau(\mu) \rightarrow +\infty.
\label{limit def}
\end{equation}
Let us use the expanding solution of the Case 3.
In accordance with the definition of Hubble function (\ref{Hubble_def}) we
calculate first $R^{\prime}(\mu,\tau)$:
\begin{eqnarray}
R^{\prime}(\mu,\tau) =
\frac{
\left(R^{\prime}_0\,k +\frac{1}{3}\,R_0\,k^{\prime}
\right)sh\,\alpha + \frac{2}{3}R_0\,k\,F^{\prime}ch\,\alpha
}{
k^{2/3} sh^{1/3}\,\alpha
}
\label{R prime}
\end{eqnarray}
where
\begin{equation}
\alpha = {\bf F}_{1}(\mu)\, + \frac{\tau}{\tau_0}
\label{alpha}
\end{equation}
In the limit of big time the Hubble function is
\begin{eqnarray}
h(\mu,\tau) = \pm \frac{2}{3 \, \tau_0}
\label{h f}
\end{eqnarray}
and does not depend on time and initial conditions,
but depends only on $\Lambda$. The density
\begin{eqnarray}
4 \pi \delta(\mu,\tau) =
\frac{1}{R^2\,R^{\prime}}
\label{rho-n.u.}
\end{eqnarray}
in the limit (\ref{limit def}) depends on time as well:
\begin{eqnarray}
\delta(\mu,\tau\rightarrow \infty) = \Phi(\mu)\,
e^{-2\frac{\tau}{\tau_0}},
\label{rho-limit}
\end{eqnarray}
where the function $\Phi(\mu)$ represents the initial conditions.

\section{The Structure of the Solution for Alternating Sign Function
$k(\mu)$}

The solutions has been built in sections (\ref{solution-1}) -
(\ref{solution-3}) describe
the three different "pure" cases of the evolution of the initial profile of
the density. These
cases are differ by the form of solution but have the common propriety:
the sign of the function $k(\mu)$ does not depend on $\mu$.
The general case of the initial conditions (\ref{init}) and (\ref{init-c})
supposes the function
$k(\mu)$ with alternating sign. This case  will be studied here.

The main idea of building the mixed solution is to satisfy the condition of
the nonintersection of laeyrs of the particles (Lagrangian coordinate is
not dependent on time).
In accordance with this idea
the mixed falling solution has the following
structure. The central part of the gaze is described by the solution "Case
3", where the particles fall during the time (\ref{zero velocity 3}) with
nonzero velocity. The shell falls to the centre during the time
(\ref{zero velocity}),
stops at time $\bar\tau$ and then goes
to the infinity (Case 1). Case 1 and Case 3 are boundaried by the case 2.
After this qualitative analysis
the building of the mixed solution is reduced to the study of the condition
of nonintersection of the layers of particles.

First of all we note that it is follows from the formulas
(\ref{case 1 peculiarity}),
(\ref{case 2 peculiarity})
and (\ref{zero velocity 3}) that the function $\dot\omega^2_0(\mu)$ should
monotonically decrease. Let us write the "pure" solutions in the form
\begin{eqnarray}
R_{i}(\mu,\tau) = (R_0)_{i}(\mu) \, r_{i}(\mu,\tau),
\label{R R r}
\end{eqnarray}
where the index ${}_{i}$ points the number of the "pure" solution: Case 1,
Case2 or Case 3.
The condition of the nonintersection has the following
form: if at the moment
of time $\tau = 0$ the Euler coordinates of three particles satisfied
the inequality
\begin{eqnarray}
R_{01}(\mu) < R_{02}(\mu) < R_{03}(\mu),
\label{condition in}
\end{eqnarray}
then this inequality holds for $\tau > 0$:
\begin{eqnarray}
R_{01}(\mu) < R_{02}(\mu) < R_{03}(\mu).
\label{condition in 1}
\end{eqnarray}
We obtain the following inequality expressing the conditions
of the nonintersection:
\begin{eqnarray}
R_{03}(\mu)\,
|k_{3}(\mu)|^{1/3}
\,
sh^{2/3}\left({\bf F}_{3}(\mu)\,\pm \frac{\tau}{\tau_0}\right),
<
R_{02}(\mu)
e^{\pm\frac{2}{3}\frac{\tau}{\tau_0}}
< \nonumber\\
R_{01}(\mu)\,
k_{1}^{1/3}(\mu)
\,
ch^{2/3}\left({\bf F}_{1}(\mu)\,\pm \frac{\tau}{\tau_0}\right).
\label{condition in 2}
\end{eqnarray}
In the limit $\tau \rightarrow +\infty$ this inequality has the form
\begin{eqnarray}
R_{03}(\mu)\,\,
|k_{3}(\mu)|^{1/3}
\,\,
e^{\frac{2}{3}{\bf F}_{3}(\mu)}
<
2^{2/3}\,\,
R_0^{2}(\mu)
<
R_{01}(\mu)\,\,
k_{1}^{1/3}(\mu)
\,\,
e^{\frac{2}{3}{\bf F}_{1}(\mu)}
\label{condition in 3}
\end{eqnarray}
and does not dependent on time, depending only on the initial conditions.
We emphasize here that all functions $R_{01}(\mu)$, $R_{01}(\mu)$,
$R_{01}(\mu)$, $k_1(\mu)$ and $k_3(\mu)$ are the fragments of the
continuous functions $R(\mu)$ and $k(\mu)$.

\section{Discation and Conclusion}

This article studies the exact solution in the LTB model, produced by the
initial conditions (\ref{init}) - (\ref{init-c}). It is shown that:
\begin{itemize}
\item the problem under consideration is defined for initial conditions
(\ref{init}) satisfied the restriction (\ref{vv});
\item in the limit of small timethe falling solution is equal in the main
part the Bonnor solution \cite{Bonnor:72} and defined by the initial
conditions;
\item in the limit of big time the solution 'forgets' the initial
conditions and has the asymptotic form (\ref{2-asimpt - >>});
\item in the lmimt of $\Lambda \rightarrow 0$ the input equation is
reduced to the equation, has been studied in \cite{A.G.2} where it is shown
the following propertys of this solution: when $\Lambda = 0$ and
$\dot\omega^2_0(\mu) = 0$ the input equation is equal to FRW model;
when $\dot\omega^2_0(\mu) \ne 0$ the FRW model is the main part of the
solution in the limit of big time;
\item the input equation is strongly depends on the sign of the function
$k(\mu)$. There are three solutions defined by the request of constant sign
of the function $k(\mu)$: "stipping of the shell" for $-1 < k(\mu) < 0$,
"collapce" for $sign k(\mu) = 0$ and "core" or "accretion" for $sign k(\mu)
> 0$;
\item the structure of the mixed solution defined by the request of the
nonintersecting of laeyrs of particles and present in the figure 2;
\item in the limit of big time the Hubble function is constant but the
density is depent on the time as $\exp{-2\frac{\tau}{\tau_0}}$;
\end{itemize}

\section{Acknowledgements}

I'm grateful to Prof.Arthur D.Chernin, Dr.Yurij Barishev, Prof. Evgenij
Edelman, Dr.Roman Zapatrin, Dr. Vadim Perepelovsky for encouragement and
discussion. I thanks Dr.Andrzej Krasinski, hi sent me his book which is
very useful in my work.
This paper was financially supported by "COSMION" Ltd., Moscow.

\end{document}